\documentclass[twocolumn]{HTL-author}

\usepackage{blindtext,graphicx}
\usepackage[absolute]{textpos}
\usepackage{multirow}
\usepackage[labelfont=bf,font=scriptsize]{caption}
\begin{document}
\begin{textblock}{17}(1.7,1)
\noindent\footnotesize This paper is a preprint of a paper submitted to IET Healthcare Technology Letters. If accepted, the copy of record will be available at the IET Digital Library.
\end{textblock} 
\vspace{5mm}
\title{Efficient Implementation of LMS Adaptive Filter based FECG Extraction on an FPGA }

\author{Bhavya Vasudeva,
        Puneesh Deora,
        P. M. Pradhan, and S. Dasgupta}

\address{Department of Electronics and Communication Engineering, Indian Institute of Technology Roorkee, Uttarakhand, India\\
E-mail: bvasudeva@ec.iitr.ac.in}


\abstract{In this paper, the field programmable gate array (FPGA) implementation of a fetal heart rate (FHR) monitoring system is presented. The system comprises of a preprocessing unit to remove various types of noise, followed by a fetal electrocardiogram (FECG) extraction unit and an FHR detection unit. In order to improve the precision and accuracy of the arithmetic operations, a floating point unit is developed. A least mean squares algorithm based adaptive filter (LMS-AF) is used for the purpose of FECG extraction. Two different architectures, namely series and parallel, are proposed for the LMS-AF, with the series architecture targeting lower utilization of hardware resources, and the parallel architecture enabling less convergence time and lower power consumption. The results show that it effectively detects the R peaks in the extracted FECG with a sensitivity of 95.74\% to 100\% and a specificity of 100\%. The parallel architecture shows upto 85.88\% reduction in the convergence time for non-invasive FECG database while the series architecture shows 27.41\% reduction in the number of flip flops used when compared with the existing FPGA implementations of various FECG extraction methods. It also shows an increase of 2 to 7.51\% in accuracy when compared to previous works.}

\maketitle

\section{Introduction}Over the past few decades, analysis of fetal electrocardiogram (FECG) has proven to be a tool of great importance when it comes to monitoring the well-being of the fetus during pregnancy and labour, unearthing vital information like fetal heart rate (FHR), heart rate variability, etc. Any abnormalities in these parameters indicate that the fetus is under distress, possibly due to asphyxia, which is a major cause of neonatal deaths. Regular FHR monitoring can enable a clinician to intervene in due time to prevent such cases.

\par Various methods [1] used to obtain FHR are auscultation (Doppler ultrasound, fetoscope), fetal phonocardiography, fetal magnetocardiography, and other invasive methods where electrodes have direct contact with fetal skin. These methods are not suitable for mobile, regular, low-cost, real-time monitoring of the fetus. An alternative method to obtain FHR is to calculate it from FECG, which can be extracted from the non-invasive abdominal ECG recordings acquired from a pregnant subject. This ECG signal contains FECG contaminated with maternal ECG (MECG), power line interference, motion artifacts, etc. Various techniques [1] involving statistical and time domain analysis have been exploited to extract the FECG. Adaptive filtering [2], nonlinear decomposition [3], blind source separation [4] (independent component analysis (ICA) based methods), approaches based on neural networks [5] are widely used.
\par Although the methods based on ICA do perform better than those which use single channel recordings, they require acquisition of multi-channel abdominal signals which may be uncomfortable for the subject. Such methods also require visual inspection of the signals, and an appropriate number of data segments have to be selected manually for a representative template [1]. The real-time implementation of such methods is not suitable unless block delayed analysis is considered [6]. Among the methods using single channel recordings, Kalman filter based approaches offer a high performance, but their implementation complexity is significantly high and they require the R peaks of the signal to have a consistent morphological shape [6]. As fetal signals and other interferences are not always linearly separable, nonlinear decomposition based methods can be used for extracting FECG in such cases. However, such methods require some prior information about the desired and undesired parts of the signal, and have a high computational complexity [6]. As the adaptive filter is an accurate method for FECG extraction and its computational complexity is relatively low [1], a least mean squares algorithm based adaptive filter (LMS-AF) is chosen for this study.
\par The signal strength of the FECG is low as compared to the MECG [6]. In rural areas, where cellular connectivity is low, the transmission of these signals for processing in cloud is not suitable. 
The FPGA can eliminate the need for an extra standby computing device that would be required for computational purposes. Also, it is a better prototyping platform for hardware implementation compared to traditional digital signal processors (DSPs). The FPGA implementation can also serve as a step towards the development of a low-cost FHR monitoring system as a system on chip. 

\par Previous hardware implementations of LMS-AF focusing on FECG extraction are discussed below. The work presented by Hatai \textit{et al.} [7] was tested on synthetic data, with upto 88\% accuracy. As synthetic data has significantly better morphology of the PQRST complex and lower noise as compared to real signals, no preprocessing was performed. Dynamic thresholding was used for peak detection. Morales \textit{et al.} [8] used a field programmable analog array for analog signal preprocessing, and an FPGA for FECG extraction with accuracy in the range of 87\% to 93\%. LabVIEW FPGA module was used to generate the hardware design.
Ortega \textit{et al.} [9] implemented the LMS-AF on a digital signal controller, used a low-noise analog front end to achieve an 93.1\% accuracy and 87.1\% sensitivity.
Some other methods for FECG extraction [10]--
[14] have also been implemented on hardware. Some of the aforementioned works [8][11][12] reportedly use fixed-point arithmetic, which leads to lower precision as compared to floating point (FP) arithmetic. It has also been reported that FP operations are difficult to implement on FPGA as the algorithm is very complex and leads to excessive consumption of logic elements [15]. 
\par The main contributions of this paper are as follows:
\begin{itemize}
    \item For the fetal R peak detection, a norm for the determination of the threshold is proposed to avoid the detection of false positives.
    \item A floating point unit (FPU) is developed for the FPGA implementation to support FP calculations, and hence improve the precision and accuracy of the system.
    Although Xilinx has a core for FP operations, it is not an open source IP, and thus cannot be used for ASIC designing, which is why the FPU is developed.
    The system is tested on both real and synthetic ECG signals, and the results validate the robustness of the system.
    \item For the implementation of the LMS-AF module, two different architectures, namely series and parallel, are proposed and compared. While the former is developed for lower hardware utilization, the latter is better in terms of lower latency and power consumption. FPGA implementation and simulation results validate the same.
\end{itemize}
\section{Methodology}
\subsection{Preprocessing}
The information that needs to be extracted from the thoracic and abdominal signals is masked by various types of noise [6], like power line interference at 50 Hz, low frequency baseline wander, broadband muscle noise, motion artifact, etc. In order to retain the MECG and FECG components [6] and attenuate the sources of noise, the signals are preprocessed.


\par In order to remove the high frequencies, a fourth order low pass Butterworth filter is used. The cutoff of the filter is kept at 45 Hz, so that the ECG components in the signal are retained [6]. For the low pass filter, the Bessel, Butterworth, Chebyshev, and resistor-capacitor (RC) filters [16] were considered. Among these, the Bessel filter offers a slower transition from pass-band to stop-band as compared to the other filters of the same order. The Chebyshev filters have ripple in the pass-band, while Butterworth and Bessel filters do not. Moreover, Butterworth filters have a significantly better frequency response (flat in the pass-band) than a simple RC filter of the same order. Therefore, a Butterworth filter is used in this work. As the cutoff of the filter is not sharp, the frequencies above 35 Hz and below 55 Hz lie in the transition band of the filter.

\par The peak at 50 Hz due to the power line interference is not sufficiently attenuated by the low pass filter. 
Therefore, a notch filter [16] centered at 50 Hz (quality factor 25) is used. 


\par Another source of noise is the baseline wander, which is a low frequency noise resulting from the respiration or movement of the subject or electrodes during recording. Since only the components corresponding to the baseline wander need to be removed, a high pass filter is not used for this application. Three techniques, namely polynomial fitting using polynomial regression, two stage median filtering and two stage moving average filtering [17] are considered to obtain an approximation of the baseline wander present in the signal. The complexity of these methods is $O(mN^2)$, $O(Nnlog(n))$, and $O(N)$, respectively where $N$ is the total number of samples, $m$ is the order of polynomial, and $n$ is the window size. Two stage moving average filter is used in this work as it is the most efficient of all and gives a smooth approximation of the baseline wander. The operations performed are summarized in the following equations:
\begin{equation}
    M_{1}[n]=\frac{1}{N_{1}}\sum_{i=0}^{N_{1}-1} x[n+i-N_1+1]
\end{equation}
\begin{equation}
    M_{2}[n]=\frac{1}{N_{2}}\sum_{j=0}^{N_{2}-1} M_{1}[n+j-N_2+1]
\end{equation}

\par where $x$ is the input signal, $M_1$ is the first stage mean with window size $N_1$, $M_2$ is the second stage mean with window size $N_2$, and $n$ is the sample index. $N_1$, $N_2$ are kept as 200 in this work. To remove baseline wander, the output of the two stage moving average filter is subsequently subtracted from the input signal. 
\subsection{LMS Algorithm}
In order to separate the FECG from the preprocessed thoracic and abdominal ECG signals, LMS-AF [18] is used. Let \textbf{x}$[n]$ = [\(x[n], x[n-1],\,.\,.\,.,x[n-m+1]\)]$^T$, represent the input to the filter, where $x[n]$ is the sample value at instant $n$, \(m\) is the order of the filter, and $(.)^{T}$ denotes transpose operator. \textbf{w}$[n]$ = [\(w_{m-1}[n], w_{m-2}[n],\,.\,.\,.,w_{0}[n]\)], is the weight vector, where $w_{m-k}[n]$ is $k^{th}$ weight at sample instant $n$. The output of the filter, at the $n^{th}$ sampling instant is given by,
\begin {equation}
 y[n]= \textbf{x}^{T}[n]\textbf{w}[n]
\end {equation}
\par The error signal is calculated as
\begin {equation}
 e[n] = d[n] - y[n]
\end {equation}
\par where \(d[n]\) is the desired signal. The weight updation is carried out as follows:
\begin{equation}
\begin{aligned}
w_{k}[n+1] &= w_{k}[n] - \mu\nabla(e^{2}[n])\\
&= w_{k}[n] + 2\mu\, e[n]\,x[n-m+k+1]
\end{aligned}
\end{equation}
\par where \(\mu\) is the step size, \(\nabla\) is the gradient operator, and \(k = 0, 1, . . . , m-1\). For this work, the thoracic signal is considered as the desired signal \(d[n]\), and the abdominal signal is the input \(\textbf{x}[n]\). The criteria for convergence of the filter weights is satisfied around 12\,000 samples. \(m\) and \(\mu\) are set to 19 and $7\times 10^{-5}$, respectively. 
\par The MECG components in thoracic and abdominal signals are not exactly same [1] which leads to some residual maternal R peaks in the resulting error signal $e[n]$ [18]. After the convergence of weights of the filter, the FECG is enhanced and the MECG is attenuated in $e[n]$ (the output of LMS-AF). 
\subsection{FHR Detection}
A modified version of the Pan and Tompkins algorithm [19] is used to detect the fetal R peaks. The output of the LMS-AF is differentiated, squared, and then passed through a mean filter of length 40. Since the extracted FECG contains residual maternal R peaks as well as sharper fetal R peaks, these operations enhance the fetal R peaks. Hence, the resultant signal $sdm$ has higher amplitude for fetal R peaks as compared to the maternal R peaks. In order to determine the threshold value $th$ which can be used to distinguish between the fetal and maternal R peaks, a new norm is proposed. The mean $m_1$ of the signal $sdm$ is used as a threshold to determine the local maxima (locations as well as amplitudes) above this threshold, present in the signal. The mean $m_2$ of the these local maxima is calculated. The threshold $th$ used to detect the fetal R peaks is set as the mean of $m_1$ and $m_2$. Among the local maxima already determined, those with amplitude less than $th$ are discarded. For the remaining local maxima, if the immediate next local maxima lies within 200 samples, the location of the local maxima with the larger amplitude of the two denotes the fetal R peak. The maximum FHR can be 200 beats per minute (bpm) [20] which corresponds to 300 samples (for a sampling frequency of 1 kHz). Therefore, maxima separated by at least 200 samples (300 bpm) are considered for detection of fetal R peaks.
\par The difference between the consecutive R peaks, as detected in the previous stage, is the RR interval. As the weights of the LMS-AF converge around 12\,000 samples, only the RR intervals for peaks occurring after 12\,000 samples are considered. The average of these RR intervals is taken, and divided by the sampling frequency to get the average RR interval length in seconds. The FHR is calculated as follows:
\begin{equation}\textrm{\textit{FHR (bpm)} = } \frac{\textrm{\textit{60}}}{\textrm{\textit{RR interval length (s)}}}\end{equation}

\section{Implementation on FPGA}For the purpose of FPGA implementation, the proposed system is divided into four units as shown in Fig. 1.
\begin{figure}[h!]
\center
\includegraphics[scale=0.19]{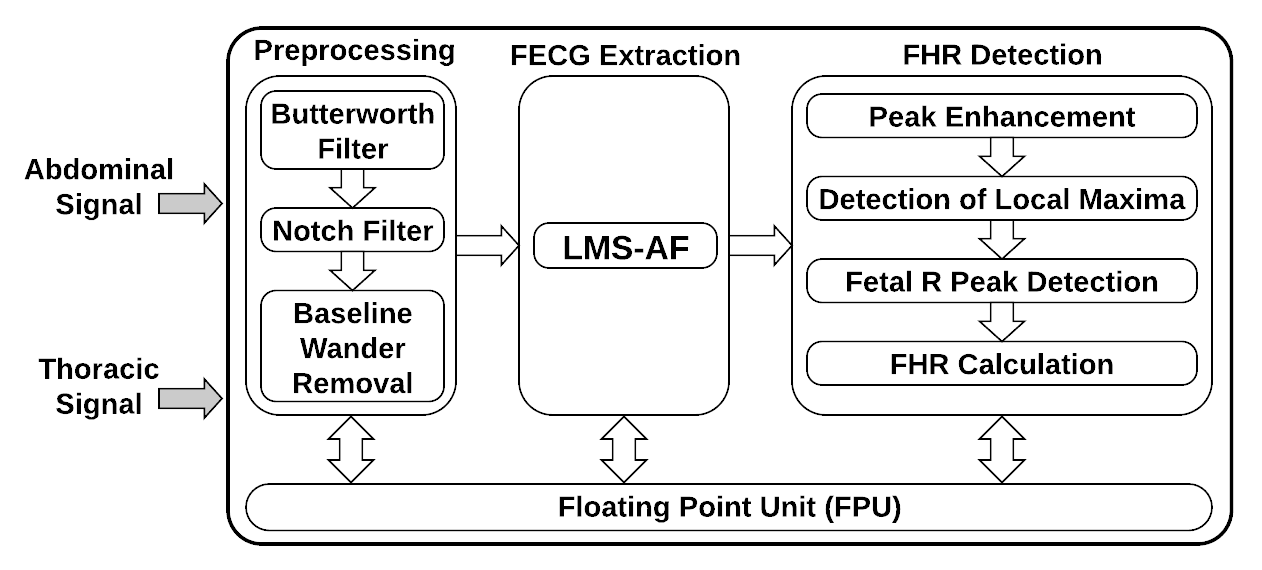}
\caption{\textit{Block diagram of the FPGA implementation of the system.}}
\end{figure}
\subsection{FPU}Since the input values to the system are FP numbers, logic cannot be defined directly on such numbers in Verilog. Therefore, an FPU is developed for performing basic arithmetic operations (addition, subtraction and multiplication) and comparison. The FP numbers are converted to their 32-bit binary representation as per the IEEE 754 standard [21], in which the first bit is the sign bit $s$, followed by 8 bits for exponent $e$, and 23 bits for fraction $f$. The value of the number is given by $(-1)^s\, 1.f\, 2^{(e-bias)}$, where $1.f$ is the mantissa $m$ and $bias$ is $2^7-1$. The inputs to the FPU module include a 2-bit sequence to select one of the four available operations and two 32-bit FP numbers ($A$ and $B$). When the numbers enter the module, they are split into three parts, i.e. sign, exponent and mantissa denoted by $s_a, e_a, m_a$ and $s_b, e_b, m_b$ for $A$ and $B$, respectively. $m$ is stored in the form of 24 bits with 1 concatenated to 23 bits of $f$. $s_{out}, e_{out},$ and $m_{out}$ denote the sign, exponent and mantissa of the output.

\par The procedure followed for the FP adder is listed below:
\vspace{-1mm}
\begin{enumerate}[i.]
\item if($e_a=e_b$) then $e_{out}=e_a$
\item else if($e_a>e_b$) then $e_{out}=e_a,\,d=e_a-e_b,\,m_b=m_b>>d$
\item else $e_{out}=e_b,\,d=e_b-e_a,\,m_a=m_a>>d$
\item end if
\item if($s_a=s_b$) then $m_{out}=m_a+m_b,\,s_{out}=s_a$
\item else
\item \hspace{5mm} if($m_a>m_b$) then $s_{out}=s_a,\,m_{out}=m_a-m_b$
\item \hspace{5mm} else $s_{out}=s_b,\,m_{out}=m_b-m_a$
\item \hspace{5mm} end if
\item end if
\end{enumerate}
\vspace{-1mm}
\par Here, $>>$ denotes the right shift operation. A similar procedure is followed for the FP subtractor, except that when the sign bits are same, subtraction is performed after comparing the mantissas and when they are opposite, addition is performed. The outputs of the FP multiplier are given by $s_{out}=s_a \oplus s_b$, $e_{out}=e_a+e_b-bias$, and $m_{out}=m_a\times m_b$, where $\oplus$ denotes the bit-wise XOR operation. For all the three operations, the next stage is to normalize the output. When $m_{out}$ is not of the form $1.f_{out}$, a repetitive process of shifting $m_{out}$ left by one place and subtracting 1 from $e_{out}$ is followed till the first bit of $m_{out}$ becomes 1. $s_{out}$, $e_{out}$ and $f_{out}$ are concatenated to get the 32-bit output of the operation.
\par For FP comparison, let $c_{out}$ denote a 2-bit sequence to denote the three cases, i.e. $A>B\, (c_{out}=01)$, $A=B\, (c_{out}=00)$, and $A<B\, (c_{out}=10)$. The procedure is listed below:
\vspace{-1mm}
\begin{enumerate}[i.]
\item if($s_a>s_b$) then $c_{out}=[10]$
\item else if($s_b>s_a$) then $c_{out}=[01]$
\item else 
\item \hspace{5mm} if($e_a>e_b$) then $c_{out}=[01]$
\item \hspace{5mm} else if($e_b>e_a$) then $c_{out}=[10]$
\item \hspace{5mm} else
\item \hspace{10mm} if($f_a>f_b$) then $c_{out}=[01]$ 
\item \hspace{10mm} else if($f_b>f_a$) then $c_{out}=[10]$
\item \hspace{10mm} else $c_{out}=[00]$ 
\item \hspace{10mm} end if
\item \hspace{5mm} end if
\item end if
\end{enumerate}
\vspace{-1mm}
\par As the output should be a 32-bit number for uniformity, thirty $0s$ are concatenated to $c_{out}$ to get the output.
\subsection{Preprocessing}
The following three modules are implemented:
\subsubsection{Butterworth Filter}In this module, one input value is used in every clock cycle to get the output as follows [16]:
\begin{equation*} \resizebox{\columnwidth}{!}{$O[k] = \alpha \,I[k] + \beta \,O[k-1] + \gamma \,O[k-2] + \delta \,O[k-3] + \epsilon \,O[k-4]$}\end{equation*}
\par where \(I[k]\) is the sample value at instant \(k\) and \(O[k]\) is the output value. The values of the constants are obtained from the transfer function of the filter, as per the cutoff and order of the filter. In this work, \(\alpha=0.00308\), \(\beta=3.28391\), \(\gamma=-4.08689\), \(\delta=2.28117\) and \(\epsilon=-0.48140\).
  \subsubsection{Notch filter}This module works in a similar manner as the previous module, following the equation [16]:
\begin{equation*}\resizebox{\columnwidth}{!}{$O[k] = \alpha \,I[k] + \beta \,I[k-1] + \gamma \,I[k-2] + \delta \,O[k-1] + \epsilon \,O[k-2]$}\end{equation*}
\par In this case, \(\alpha=0.99405\), \(\beta=-1.31278\), \(\gamma=0.99405\), \(\delta=1.31272\) and \(\epsilon=-0.98804\).
\subsubsection{Baseline wander removal}Fig. 2 shows the structure of the two stage moving average filter. As in $(1)$, the first stage mean $M_1$ is the average of $N_1$ values. In every clock cycle, the input is added to $M_1$ and $x[N_1-1]$ is subtracted from $M_1$, both after getting multiplied by $\frac{1}{N_1}$. For the moving average operation, all the values in Memory 1 are shifted by one position, so that $x[N_1-1]$ is discarded and a new value is stored in $x[0]$. A similar procedure is followed for calculating the second stage mean $M_2$ as per $(2)$. $M_1$ is multiplied by $\frac{1}{N_2}$, stored in Memory 2 and also added to $M_2$. The last value of Memory 2 can then be directly subtracted from $M_2$ to obtain the second stage mean. The values of $M_2$ represent the baseline wander approximation. The output of the two stage moving average filter is used to remove the baseline wander from the input by performing one subtraction operation in every clock cycle. The latency of each of these modules is 1 clock cycle.
\begin{figure}[h!]
\center
\includegraphics[scale=0.28]{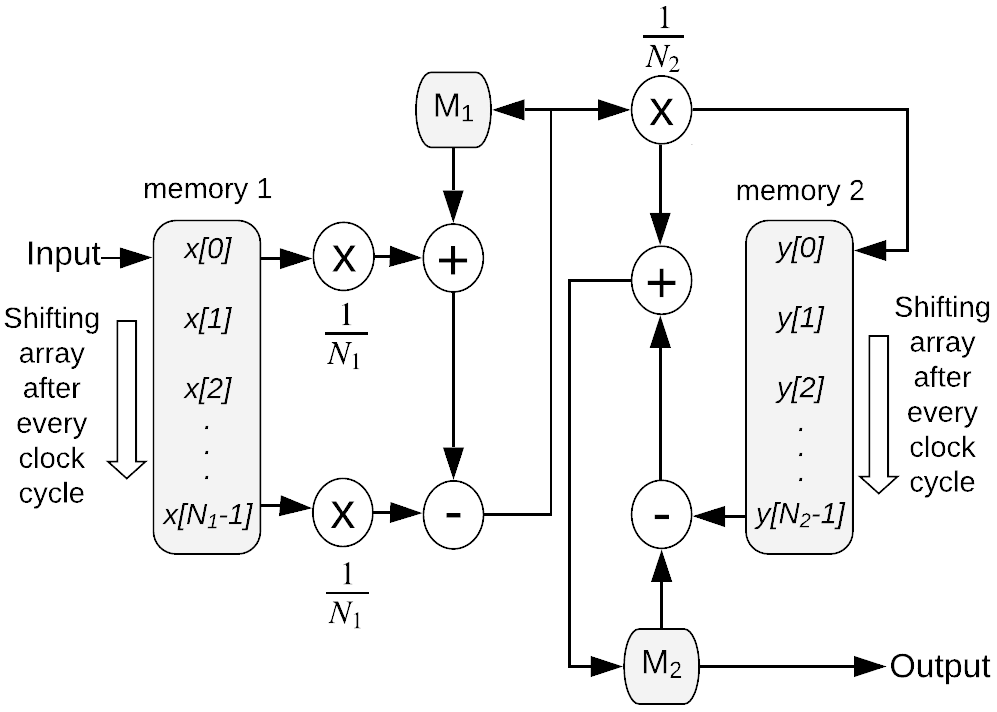}
\caption{\textit{Structure of the two stage moving average filter.}}
\end{figure}
\subsection{FECG Extraction}
This stage comprises of the LMS-AF, for which two different architectures are proposed. Figs. 3(a) and (b) illustrate the proposed series and parallel architectures of LMS-AF module, respectively. In both the figures, $\beta=2\mu$. As the magnitude of abdominal and thoracic signals may not be of the same order, they have to be scaled appropriately (denoted by \textit{scaling} in the figures) before being used in the algorithm.

\subsubsection{Series Architecture} In Fig. 3(a), Memory 1 stores the vector \(\textbf{x}^T[n]\) and an extra element, and Memory 2 contains the weights of the filter. In every clock cycle, one element of \(\textbf{x}^T[n]\) is scaled, multiplied by one element of \(\textbf{w}[n]\) and added to \(y[n]\). The same element of $\textbf{x}^T[n]$ is also copied to the immediate next position in $\textbf{x}^T[n]$. Thus, after \(m\) clock cycles, \(y[n]\) has been obtained as in $(3)$, and \(\textbf{x}^T[n]\) has shifted by one index. In the following clock cycle, error is calculated using $(4)$, and the updated value of the first weight of the filter is also obtained. This updated weight value is stored in its position in the next clock cycle. This sequential process is repeated until all the weights are updated, which corresponds to \(m+1\) clock cycles. After a total of \(2m+1\) clock cycles, a new input value is stored in $x[0]$ so that $\textbf{x}^T[n]$ is updated. The register containing \(d[n]\) also gets updated. As the required output for a particular pair of $\textbf{x}^T[n]$ and $d[n]$ is obtained after $2m+1$ clock cycles, the latency of this module is $2m+1$ clock cycles.
\subsubsection{Parallel Architecture} In Fig. 3(b), the Memory 1 (vector \(\textbf{x}^T[n]\)) gets updated with the next input value in every clock cycle. Each element of $\textbf{x}^T[n]$ is scaled, and then multiplied with the elements from the Memory 2 (vector \(\textbf{w}[n]\)). These are added to obtain \(y[n]\), as in $(3)$. The register containing \(d[n]\) is updated in every clock cycle, and is used to calculate the error, using $(4)$. Since \(2\mu e[n]\) is used in every weight updation, it is calculated first, and subsequently multiplied with the values from Memory 1 to update the weights, using $(5)$. The updated weights are stored in Memory 2. Thus, all the operations involving the error calculation and weight updation are performed in a single clock cycle and the latency of this module is 1 clock cycle.
\begin{figure}[h!]
\center
\includegraphics[scale=0.188]{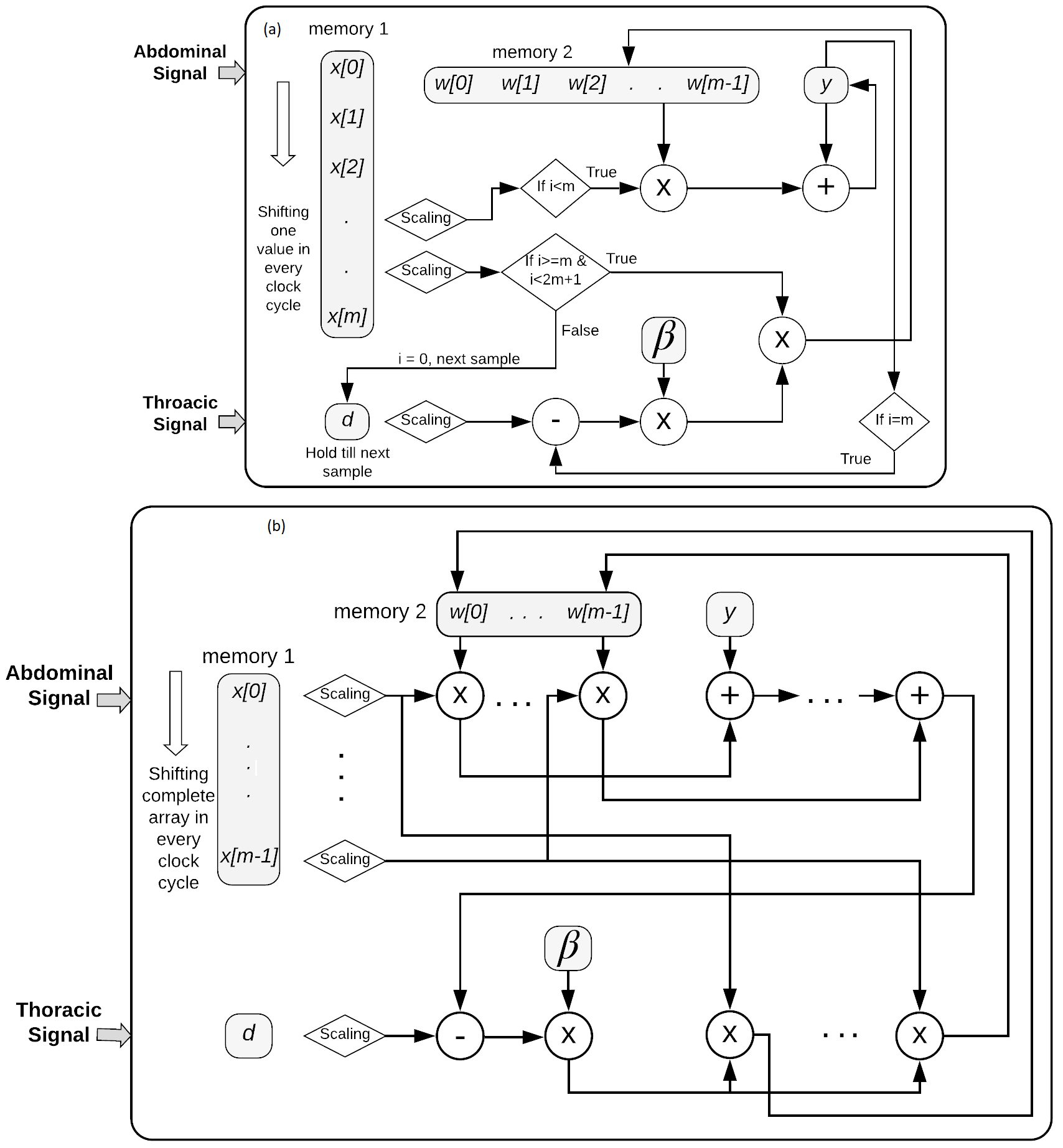}
\caption{\textit{Illustration of proposed (a) series and (b) parallel architecture of LMS-AF.}}
\label{fig:my_label}
\end{figure}

\subsection{FHR Detection}
The four modules of this stage are:
\subsubsection{Peak Enhancement}In this module, the input is differentiated, squared, and passed through a mean filter of length $P$. The mean $m_1$ of $sdm$, which is required in the next module, is also determined in this module. 
The operations executed in every clock cycle are summarized below:
\begin{enumerate}[i.]
\item \(pval=cval\)
\item \(cval=input\)
\item \(sdiff=(cval-pval)\times(cval-pval)\)
\item \(M[0]=sdiff\times \frac{1}{P}\)
\item \(sdm=sdm+M[0]-M[P-1]\)
\item $m_1=m_1+sdm\times \frac{1}{N}$
\item \textit{Shift the elements of $M$ by one position}
\end{enumerate}
\par Here, $cval$ and $pval$ denote the current and previous input values, respectively. $sdiff$ denotes the differentiated and squared signal, which is stored in the memory $M$ (size $P$) after multiplication by $\frac{1}{P}$. $N$ denotes the number of input samples.
\subsubsection{Detection of Local Maxima}In this module, the local maxima are determined, using $m_1$ as threshold. The operations executed in every clock cycle are summarized below:
\begin{enumerate}[i.]
    \item if($in>m_1$ and $in>R_1$) then $R_3=in, R_4=R_2$
    \item else if ($in<m_1$) then $pv=R_3, pl=R_4, m_2=m_2+\frac{pv}{N}$
    \item end if
    \item $R_1=in$
    \item $R_2=R_2+1$
    \item if($R_2=N$) then $th=\frac{m_1+m_2}{2}$
    \item end if
\end{enumerate}
\par Here, $in$ denotes the current input, $R_1$ ($R_2$) is used to store the input value (location) for the next clock cycle, and $R_3$ ($R_4$) is used to conditionally store the input value (location). The locations and values of local maxima are denoted by $pl$ and $pv$, respectively, $m_1$ and $m_2$ are initialized to zero.

\subsubsection{Fetal R Peak Detection}In the first cycle, the inputs $pl$ and $pv$ are stored in $R_1$ and $R_2$, respectively. The operations executed in every clock cycle are summarized below:
\begin{enumerate}[i.]
    \item if($pv>th$) then
    \item \hspace{5mm} if($pl-R_1>200$) then $out=R_1, R_1=pl, R_2=pv$ 
    \item \hspace{5mm} else
    \item \hspace{10mm} if($pv>R_2$) then $out=pl, R_1=pl, R_2=pv$
    \item \hspace{10mm} else $out=R_1$
    \item \hspace{10mm} end if
    \item \hspace{5mm} end if
    \item end if
\end{enumerate}
\par Here, $out$ denotes the locations of the fetal R peaks detected.
\subsubsection{FHR Calculation}The RR intervals are estimated using the differences between consecutive peak locations ($out$). Two registers are used for storing the current input and the previous input. The estimated RR intervals are accumulated and averaged out, after which FHR is obtained using $(6)$.

\section{Results and Discussion}
To test the system for real signals, the non-invasive FECG (NiFECG) database [22] and Database for Identification of Systems (DaISy) [23] are used. In the NiFECG database the signals have been sampled at 1 kHz with 16-bit resolution. From the dataset to be tested, 1 thoracic and 1 abdominal signal are chosen as inputs to the system. These are shown in Figs. 4(a) and (b), respectively. The DaISy database consists of 8 channel, 10 sec recordings sampled at 250 Hz, where 3 channels are the thoracic signals and the remaining 5 are the abdominal signals. The synthetic signals were simulated using FECGSYN toolbox [24] in MATLAB, at a sampling rate of 1 kHz. Figs. 4(c) and (d) show the thoracic and abdominal input signals. It is observed that the synthetic signals are less noisy as compared to the real signals. In the thoracic signals, the peaks correspond to maternal R peaks. In the abdominal signals, the higher peaks correspond to maternal R peaks, and the ones annotated as \textit{fpk} are fetal R peaks.
\par Figs. 5(a) and (d) show the time series for real and synthetic signals after preprocessing. The output of the LMS-AF stage, where FECG is enhanced and MECG is attenuated, is shown in Figs. 5(b) and (e). The signal obtained after peak enhancement (labelled $sdm$) and the detected fetal R peaks (denoted by $fpk$), are shown in Figs. 5(c) and (f). Table 1 lists the results obtained for sensitivity, specificity, accuracy, and FHR for the tested datasets. It is observed that the proposed norm for the determination of the threshold, which is used for fetal R peak detection, results in no false positives. The application of Pan and Tompkins algorithm [19] on ecga444 [22] dataset gives a sensitivity of 78.72\%, specificity of 48.28\% and accuracy of 67.11\%, as this method was developed for detecting R peaks in ECG signals of a single subject, whereas the output of LMS-AF has enhanced FECG as well as residual MECG.
\par Table 2 summarizes the comparison of performance of the proposed work with various FECG extraction methods. The proposed work shows an increase of 1.34\% in the sensitivity, and an improvement of 2\% in accuracy for DaISy. The proposed method also shows an increase of 1.02\% in the senstivity and an improvement of 7.51\% in the accuracy when compared to works that have tested their systems on both NiFECG and DaISy.
\begin{figure}[h!]
\center
\includegraphics[scale=0.58]{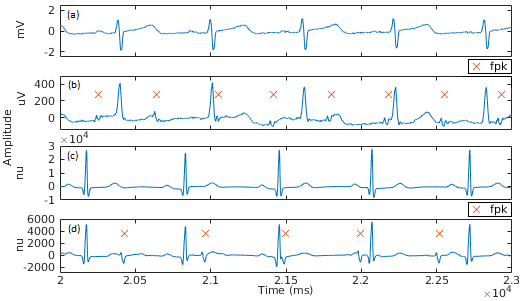}
\caption{\textit{Real (a) thoracic, and (b) abdominal signals. Synthetic (c) thoracic, and (d) abdominal signals. Synthetic dataset has no units (\textit{nu}).}}
\label{fig:my_label}
\end{figure}

\begin{figure}[h!]
\center
\includegraphics[scale=0.58]{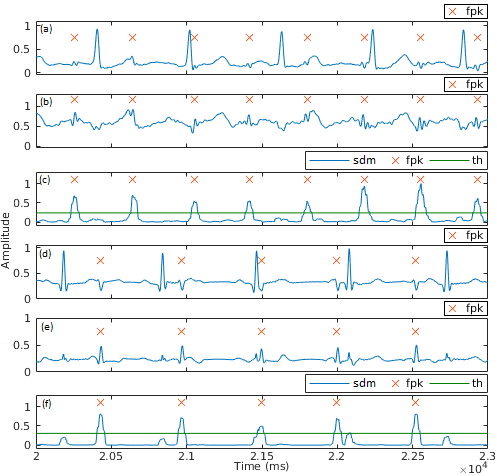}
\caption{\textit{Results of (a) preprocessing, (b) LMS-AF, and (c) peak detection for real signals, and (d) preprocessing, (e) LMS-AF, and (f) peak detection for synthetic signals. All values are normalized between 0 and 1.}}
\label{fig:my_label}
\end{figure}
\begin{table}[h!]
\centering
\caption[ghg]{\scriptsize{Results obtained for different datasets using the proposed approach.}}
\resizebox{0.95\columnwidth}{!}{
\begin{tabular}{c c c c c} 
 \hline
 \rule{0pt}{9pt}Dataset & FHR (bpm) & Sensitivity & Specificity & Accuracy \\ 
 \hline
 \rule{0pt}{9pt} ecgca444 [22] & 152 & 95.74\% & 100\% & 97.37\% \\
 \rule{0pt}{9pt} ecgca840 [22] & 161 & 96\% & 100\% & 97.37\% \\
 \rule{0pt}{9pt} ecgca746 [22] & 147 & 97.78\% & 100\% & 98.53\% \\
 \rule{0pt}{9pt} ecgca771 [22] & 153 & 100\% & 100\% & 100\% \\
 \rule{0pt}{9pt} DaISy Channel 2 [23] & 143 & 100\% & 100\% & 100\% \\  
 \rule{0pt}{9pt} DaISy Channel 3 [23] & 143 & 100\% & 100\% & 100\% \\ 
\rule{0pt}{9pt} Synthetic [24] & 115 & 100\% & 100\% & 100\% \\  
 \hline
\end{tabular}}
\\
\end{table}
\par The system is implemented on the Xilinx Artix-7 FPGA (XC7A100TCSG324-1), equipped with 63\,400 Look-Up Tables (LUTs), 126\,800 Flip Flops (FFs), 240 DSPs, and 210 I/O ports. Except for the baseline wander removal, the detection of local maxima module, and the LMS-AF module, all other modules have minimal resource ($\sim$0 LUTs and FFs) and power utilization (0.068W). The baseline wander removal module consumes 2.691W power, and utilizes 820 LUTs and 94 FFs. The power per cycle is 89.683 $\mu$W. The detection of local maxima module consumes 0.167W power, and utilizes 45 LUTs and 34 FFs. The power per cycle is 9.278 $\mu$W.
\par For the LMS-AF module, the resource utilization and power consumption depend on the architecture and the order of the filter. There is a trade-off between the series and parallel architectures in terms of the convergence time and the resource utilization. For the parallel design, the number of operations in every clock cycle is more as compared to the series design, and hence the resource utilization is greater. On the other hand, the series architecture distributes the same number of operations across more clock cycles, and hence needs more time for convergence, and consumes more power. Also, an increase in the filter order results in an increase in the number of operations as well as the resource utilization.
\begin{table}[h!]
\centering
\caption[h]{\scriptsize{Comparison of performance of proposed method with various FECG\\\centering{extraction methods.}}}
\label{table:1}
\resizebox{0.95\columnwidth}{!}{
\begin{tabular}{c c c c c} 
 \hline
 \rule{0pt}{9pt}Method & Dataset & Sensitivity &  Accuracy \\ 
 \hline
 \rule{0pt}{9pt} Tai Le \textit{et al.} [25] & DaISy & 98.68\% & 98.04\% \\
 \rule{0pt}{9pt} Gini \textit{et al.}[26] & DaISy & 91\% & 87.30\% \\
 \rule{0pt}{9pt} Lima-Herrera \textit{et al.} [27] & DaISy and  NiFECG & 97.50\% & 92.10\% \\
 \rule{0pt}{9pt} Morales \textit{et al.} [8] & DaISy and  NiFECG & - & 89\% \\
 \rule{0pt}{9pt} Proposed method & DaISy & 100\% & 100\% \\
\rule{0pt}{9pt} Proposed method & DaISy and  NiFECG & 98.5\% & 99.04\% \\  
 \hline
\end{tabular}}
\\
\begin{scriptsize}
\vspace{1mm}
- Not reported.
\end{scriptsize}
\end{table}
\begin{table}[h!]
\centering
\caption[h]{\scriptsize{Comparison of hardware implementations of proposed method and\\ various FECG extraction methods.} 
}
\label{table:1}
\resizebox{0.98\columnwidth}{!}{
\begin{tabular}{c c c c c c} 
 \hline
 \rule{0pt}{9pt}& & Convergence & Power & & \\
 Method & Device & Time & Consumption & LUTs & FFs \\
  & & (ms) & (W) & & \\
   \hline
 \rule{0pt}{9pt} LMS [7] & XC6SLX45-3-CSG394 & - & - & 1042 & 440\\ 
\rule{0pt}{9pt} LMS [8] & Spartan3E XC3S500E & 600 & - & - & -\\
 \rule{0pt}{9pt}LMS [9] & dsPIC30F6014A & 0.33 & 1.67* & - & -\\
\rule{0pt}{9pt} OL-JADE [10]& OMAP L137 & 948 & - & - & -\\
\rule{0pt}{7pt} \multirow{2}{*}{Infomax [11]} & Stratix-V & \multirow{2}{*}{3.4-54} & \multirow{2}{*}{0.55} & - & -\\  
\rule{0pt}{7pt} & 5SGXEA7N2F45C2 & & & & \\
\rule{0pt}{7pt} Neural & Stratix-II & \multirow{2}{*}{-} & \multirow{2}{*}{-} & \multirow{2}{*}{9726} & \multirow{2}{*}{4324}\\
\rule{0pt}{7pt} Network [12] & EP2S15F484C3 & & & & \\
 \rule{0pt}{9pt} BSS [13] & Spartan-3 & - & - & 3002 & 405\\
 \rule{0pt}{9pt} Proposed Series & Artix-7 & 18.72 & 6.478 & 2368 & 294\\
 \rule{0pt}{9pt}Proposed Parallel & XC7A100TCSG324-1 & 0.48 & 1.954 & 22\,407 & 640 \\
 \hline
\end{tabular}}
\\
\begin{scriptsize}
\vspace{1mm}
- Not reported. * The system proposed by Ortega \textit{et al.} [9] consumes 1W, for the current absorption of 200 mA and supply of 5 V, at 30 MHz operating frequency.
\end{scriptsize}
\end{table}
\par Table 3 summarizes the comparison between the existing implementations of various FECG extraction methods on different hardware platforms and the proposed architectures of LMS-AF after mapping the power consumption and convergence time to operating frequency of 50 MHz. As the number of cycles is different for the two architectures, there is a large difference in the values of power consumption. The power per cycle is 7.823 $\mu$W for series and 65.133 $\mu$W for parallel architecture (considering convergence at 12\,000 samples and total number of input samples as 30\, 000). The series architecture uses 9 instances of the FPU module, and shows 27.41\% reduction in the number of FFs, whereas the number of LUTs is comparable to the other methods. The parallel architecture uses 98 instances of the FPU module, and shows upto 85.88\% reduction in the convergence time when compared with the methods [8][11][13] using NiFECG database. Since $m$ is 19 in this work, and the latency is $2m+1$ clock cycles for series and 1 clock cycle for parallel architecture, the convergence time for the former is 39 times the convergence time for the latter.

\par The use of FP operations in the implementation of the proposed architectures greatly enhances the precision and accuracy of the system. Many of the methods listed in Table 3 have reportedly used fixed-point numbers. The use of fixed-point numbers would have resulted in a lower resource utilization and power consumption as the operations involving FP numbers are computationally intensive [7][13][15]. However, the use of fixed-point numbers compromises with the accuracy of the system. 
\section{Conclusion}
In this paper, the FPGA implementation of a complete system for preprocessing ECG signals, extracting FECG, and subsequently calculating FHR is presented. For the removal of high frequency components, power line interference, and baseline wander, Butterworth, Notch, and two stage moving average filters are used, respectively. For FECG extraction, an LMS-AF is used, and series and parallel architectures are designed for its implementation. The precision and accuracy of the complete system is significantly enhanced by the use of FP arithmetic, for which an FPU is developed. Comparisons with previous work show that the proposed parallel architecture requires the least time for convergence of filter weights, while the proposed series architecture has low resource utilization.

\end{document}